# A Comparative Study of Location Management Schemes: Challenges and Guidelines

Vijendra Singh Bhadauria
Department of Computer Science and Engineering
Sagar Institute of Research and Technology Excellence
Bhopal, India
mic_cs_bpl@yahoo.com

Sanjeev Sharma
Department of Information Technology
University Institute of Technology, R.G.T.U.
Bhopal, India
sanjeev@rgtu.net

Ravindra Patel
Department of Computer Applications
University Institute of Technology, R.G.T.U.
Bhopal, India
ravindra@rgtu.net

*Abstract*—One of the key issues in mobile communication is to find the current location of mobile terminal (MT) to deliver the services, which is called as location management (LM). Increasing users and diverse services demand for a high-quality skeleton for LM. As an MT moves within a cellular network, it registers its new location to the nearest base station (BS). When a call arrives for an MT, the network searches the target MT in the area where it was last registered. This paper presents comprehensive classification of existing major LM schemes, their comparative study and factors influencing their performance. Finally, guidelines for developing and rating a LM scheme are suggested with the help of LPCIC rule, which is the main contribution of this paper.

*Keywords- mobility, location management, cellular mobile networks, mobile terminal*

I. INTRODUCTION TO MOBILE COMMUNICATION NETWORK

Mobile communication systems enable MTs (subscribed mobile users) to transfer and receive various types of data between any desired locations. As shown in Fig.1, the service area is divided into location areas (LAs) and the LA is divided into cells. In each cell, there is a base station (BS) that communicates with MTs over pre-assigned radio frequencies. Groups of several cells are connected to a mobile switching center (MSC) through which, the calls are routed to telephone networks. MSC is a telephone exchange specially assembled for mobile applications. It interfaces between mobile phones (via BS) and the public switched telephone network (PSTN) or public switched data network (PSDN) making the mobile services widely accessible. Home Location Register (HLR) and the Visitor Location Register (VLR) databases are used for LM in mobile networks. The HLR contains permanent data like directory number, profile information, current location, and validation period of the MTs whose primary subscription is within the area. For each MT, it contains a pointer to the VLR to assist routing incoming calls.

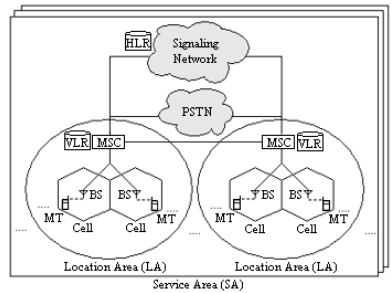

Figure 1. Location management architecture in 2-G network

A VLR is associated with each MSC that contains temporary record for all MTs currently active within the LA of the MSC. VLR retrieves the information for handling calls to or from a visiting MT. To facilitate the tracking of a moving MT, network is partitioned into many location areas (LAs). Each LA may include tens or hundreds of cells depending upon the location management scheme used. Each LA is served by a VLR. The service area served by an HLR is referred to as Service Area (SA). The network comprises of several such SAs.

The rest of the paper is organized as follows- Section-2 describes basic operations in LM. Section-3 tells about mobility models to evaluate the performance of LM schemes. Section-4 discusses about location area configurations. Section-5 and 6 present a comprehensive classification and comparative study of major location update and paging schemes. Section-7 deals with radio location schemes. Section-8 suggests the guidelines for development of a LM scheme. Section-9 gives an overview of LM architecture in 3G-networks. Section-10 concludes the paper.

## II. BASIC OPERATIONS IN LOCATION MANAGEMENT

Location management involves tracking of MT's location, moving from place to place so as to provide them services timely. Two basic operations in mobility tracking are: location update and paging. Basically, whenever MT moves out of its current LA, its geographical location information is updated to the nearest BS. On a call arrival, the network searches the called MT by sending polling signals to the vicinity of last reported location of MT. This searching process is called paging [13]. The total LM cost is generally calculated by summing up the cost of location update (LU) and paging. Normally, the LU costs higher than paging. The network can require more frequent LUs, in order to reduce paging cost. Inversely, the network may require rare LUs, storing less information about user mobility to reduce computational overhead, but at a higher paging cost. To reduce the total location management cost, it is essential to provide good trade-off among paging and LU operations [5].

## III. MODELING USER MOBILITY

Mobility pattern plays an important role in evaluating the performance of LM schemes. A mobility model is used to describe the mobility pattern of MTs; the simplest model is Random-walk model, where MT-movements are assumed to be entirely random. Each neighboring cell may be visited with equal probability. However, practically MT does not move in fully random fashion. Fluid-flow is a very common model, where rather than individual users the network is considered as a whole. The fluid flow model is suitable for vehicle traffic in highways, but not suitable for pedestrian movements with stop-and-go interruption. This model provides no insight on a smaller scale, nor does it give any prediction of specific user movements. In Markovian-mobility model, the next movements of MT are predicted on the basis of its past movements. At large computational cost, every inter-cell movement probability is defined for each user. An extension of the Markovian model, created at perhaps even greater cost, is the Activity-based model. In this model, parameters such as time of day, current location, and predicted destination are also stored and evaluated to create movement probabilities. The Shortest Path model assumes that within the LA, an MT will follow the shortest path measured in terms of number of cells traversed, from source to destination. At each cell intersection, the mobile station makes a decision to proceed to any of the neighboring cells such that the shortest distance assumption is maintained. In Selective-prediction model, predictions are only made in regions where movements are easily foreseeable, and a random prediction method is used elsewhere. In fact, research on all current models shows that none of them truly does a satisfactory job in predicting user movements, indicating a call for further research in this area.

## IV. LOCATION AREA CONFIGURATIONS

The location area can be viewed as the biggest possible area that may need to be paged while locating an MT to deliver a call. The location area can be classified as static and dynamic. Static LAs are fixed group of cells such that any MT moving through them follows same set of location management rules. It is the easiest solution to physically divide a network without any customization, but users are not homogeneous. The biggest problem in static LAs is that they undergo ping-pong effect if MT repetitively moves between two or more adjacent LAs.

Unlike static LAs, dynamic LAs may take different shape and size in order to be optimal for individual user, constructed such that the users will change their LA as rarely as possible. This minimizes the possibility of paging whole LA in search for an MT thus reduces the total paging cost up to a large extent.

## V. LOCATION UPDATE SCHEMES

Location update schemes can broadly be classified into static and dynamic. Presently, most location update schemes are static due to their simplicity. As shown in Fig.2, Static LU schemes can be classified as Always-update, Never-update, Static Interval-based, Reporting center and Paging cell. In always update, MT updates its location upon every inter-cell movement, which incurs large number of updates for highly mobile users. Under never-update, adjacent cells in the network are grouped to function as a larger single cell called location area; update is triggered only on inter-LA movement but doing so, paging cost rises substantially for MTs with high incoming call frequency. The static interval-based scheme requires MT to update its location after every predefined, uniform time period that provides a balance between the extremes of previous two schemes. Locating highly mobile users becomes difficult under this scheme. Inversely, a stationary user regularly enforces unwanted LUs.

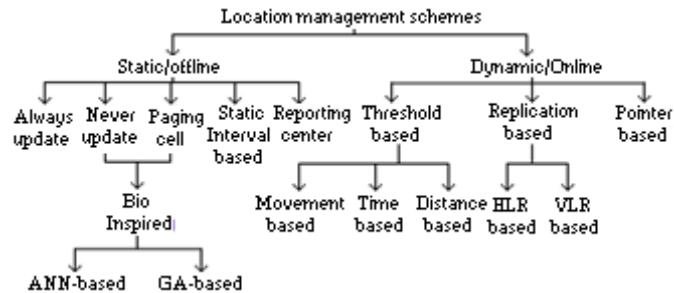

Figure 2. Location-update schemes

In reporting-center scheme, some cells are selected as reporting-centers; MT updates its location only when it enters into a new reporting-center. When a call arrives, the vicinity of the last reported reporting-center is paged to locate the target MT, selecting the reporting centers that provide quite good trade-off between paging and update operations, itself is a complex task. Paging-cell (PC) scheme groups the cells into paging areas, all cells within a paging area are paged simultaneously whereas each paging area can be paged sequentially until the target MT is located by the network upon arrival of a call, however long delay may occur in locating the MT. PC and LA based schemes are usually employed as the network topology for bio-inspired schemes that are generally based on artificial neural network or genetic algorithm. Table I illustrates the comparison between all the static LU schemes.

TABLE I. COMPARISON BETWEEN MAJOR STATIC LU SCHEMES

| LU Scheme | Update cost | Paging cost | Location accuracy | Major drawback |
|---|---|---|---|---|
| *Always update* | high | low | 1 cell | number of updates is too high |
| *Never update* | low | high | 1 location area | whole LA needs to be paged |
| *Paging cell* | low | high | several cells | long time delay in large networks |
| *Static interval based* | constant | high | several cells | unnecessary updates by stationary users |
| *Reporting center* | low | high | several cells | high computational overhead |
| *Bio inspired* | sub-optimal | sub-optimal | several cells | too much computational overhead |

Dynamic LU schemes work on the principle that the MTs follow different movement patterns therefore LU scheme should treat them differently, such that each MT has its own optimal LU standard based on its movement habits. Under threshold based schemes, LU occurs each time a parameter (time, movement or distance) goes beyond a threshold value, which can be modified on per user basis. In time-based scheme, MT updates its location at constant time intervals, which saves computation but incurs needless LUs when MT does not move. In movement-based scheme, LU takes place each time the MT crosses a predefined number of cells. It works better than time-based scheme, unless the MT is highly mobile. In distance-based scheme, MT performs LU when its distance from the cell where it performed the last LU, exceeds a predefined value. It requires MT to track distances, which increases computational overhead and so is the limited battery consumption on MT side. HLR-based replication scheme lets the LM messages to be routed to current (nearest) HLR rather than the master HLR but it imposes extra load on current HLR, thus reducing its performance. In VLR-based replication scheme, location information of MT is replicated among VLRs so that the mobility information is more readily available to the network. Its major problem is to keep these replicas consistent and up-to-date, they must be updated whenever the user profile is updated. In pointer-forwarding schemes, some updates to the HLR are avoided by setting up a forwarding pointer from the previous VLR to the new VLR. However, the penalty is the time delay for tracking current location of MT. Table II illustrates the comparison between various dynamic location update schemes.

TABLE II. COMPARISON BETWEEN MAJOR DYNAMIC LU SCHEMES

| LU Scheme | Update cost | Paging cost | Location accuracy | Major drawback |
|---|---|---|---|---|
| *Time based* | low | high | several cells | • unnecessary updates by stationary users |
| *Movement based* | low | high | several cells | • may suffer from ping-pong effect |
| *Distance based* | low | high | several cells | • high computational overhead on MT side |
| *HLR level replication* | low | high | 1 cell to several cells | • extra burden on current HLR<br>• increased call establishment delay |
| *VLR level replication* | low | high | several cells | • overhead of regularly updating distributed mobility information |
| *Pointer based* | low | high | several cells | • increased call establishment delay |

## VI. PAGING SCHEMES

In attempt to locate target MT as quickly as possible, multiple methods of paging have been proposed by the researchers. The most basic method used is Simultaneous paging, where every cell in the MT's LA is paged at the same time. If LA contains large number of cells, this scheme costs terribly high. In Sequential Paging, cells within an LA are paged one after the other, in order of decreasing user dwelling possibility. If the user resides in an infrequently occupied location, long delay may occur in finding the MT. Intelligent-Paging (optimized version of Sequential Paging) calculates the specific paging areas to poll sequentially, based upon a dwelling probability matrix. However, this scheme has too much computational overhead incurred through updating and maintaining the matrix. Rule-based paging scheme is a knowledge based approach, where the current interaction of the MT with the network is represented as a set of facts, which is used to predict the location of MT when a call arrives. Table III illustrates the comparison between various paging schemes.

TABLE III. COMPARISON BETWEEN MAJOR PAGING SCHEMES

| Paging Scheme | Paging Area | Time delay | Paging cost | Major drawback |
|---|---|---|---|---|
| *Simultaneous* | whole LA | low | high | • excessive cost for bigger LAs |
| *Sequential* | 1cell to 1LA | high | depends upon number of paging-miss | • large delay occurs if user resides in an infrequently occupied location |
| *Intelligent* | 1cell to 1LA | high | depends upon number of paging-miss | • too much computational overhead<br>• long delay occurs if user resides in an infrequently occupied location |
| *Rule based* | 1cell to 1LA | high | depends upon number of paging-miss | • too much computational overhead<br>• long delay occurs if user resides in an infrequently occupied location |

## VII. RADIO-LOCATION SCHEMES

The radio-location systems work by measuring radio signals traveling between a fixed station and the MT. The location of MT is determined, using the geometrical relationship between the fixed station and measuring the direction and/or time delay of radio signals.

Radio-location schemes can be self-positioning or remote-positioning. The former utilizes a positioning receiver in the MT through which, MT can perform signal measurements and calculate its position whereas in later, transmitter is employed in MT and receivers at some other locations that determine the position of MT after measuring the signals received from it. All these measurements are collected at a central site where the position of MT is estimated. The position of MT based on accuracy, can be determined by using various schemes such as: Cell-identification (CI), Angle of arrival (AoA), Time of arrival (ToA), Enhanced observed Time difference (E-TOD) and Assisted GPS (A-GPS).

CI is the positioning scheme that is supported by all handsets. It uses the network base transceiver station (BTS) to identify the user in the cell area. Successful operation requires that all signals arrive at the base station at the appropriate time. The MTs make measurements on the air interface and send them to the network for decisions, these measurement reports contain the estimated power level at the MT, which is used to estimate the distance between the base station and the MT. In AoA, the angles are calculated at which a signal arrives at two base stations from a MT using triangulation with complex antenna array at each cell site to find the location. Simple geometric relationships are then used to determine the location by finding the intersections of the lines-of-position, if at least two base stations are able to determine the AoA of the signal. ToA scheme measures time of arrival of the signal from the mobile user to a number of BSs located at quite accurately known positions. Usually, the clock of the mobile station is not synchronized with the network stations. This scheme requires BSs and MT to be synchronized. The large number of location measurement units (LMUs) leads the implementation of this scheme is very expansive as compared to performance enhancements it offers. E-OTD is modified version of ToA, where MT with the help of E-OTD software, measures the differences of arrival time of signals transmitted from minimum three synchronized base stations. Timing measurements made by MT are transferred to the Serving Mobile Location Center (SMLC), which calculates the distance of that MT from each of those BTS and returns it back to the MT. These time differences are combined to produce intersecting hyperbolic lines from which the location is estimated[7]. E-OTD requires considerable network investment and also requires specific software to be installed within the MT side. A-GPS is advanced positioning scheme combining mobile technology and GPS. It is expensive for the end-users, as they have to invest in a GPS-equipped MT. Adding GPS functionality has a high impact on the MT with new hardware and software required. GPS receiver needs to be in sight of four or more satellites to estimate a three-dimensional position.

## VIII. GUIDELINES FOR THE DEVELOPMENT OF LM SCHEME *(The LPCIC Rule)*

There are many parameters that influence the total LM cost[18] such as cost of database management in LU operation, cost in terms of wired line (backbone) network bandwidth used (that connects base stations to each other). Mostly these costs are assumed to be constant for all LM schemes. We suggest that the objective while developing a LM scheme should be, to minimize one or more parameters mentioned below without maximizing the other(s):

A. *Location update (LU) cost:* It is believed that the cost of a LU operation is ten times more than the cost of a paging operation. The update cost can be reduced by minimizing any one of the following

  1) Resources required in update operation

  2) Frequency of update operations and

  3) Number of messages interchange in update operation, which may vary from 5 to 36.

B. *Paging cost:* Although the cost of a LU operation is significantly greater than the cost of a paging operation[18] but paging cost can not be overlooked. It can be reduced by minimizing the area to be paged when an incoming call arrives. This can be done by making latest information (or accurate prediction) about the current location of the MT easily available to the network.

  When a call arrives for an MT then generally those cells are paged first where the target MT is most likely to be found. If not found (paging-miss occurs), then other cells are paged in succession on the basis of possibility of finding it until the MT is found (paging-hit occurs). Thus, higher number of paging-miss also leads to higher paging cost.

C. *Call establishment delay:* The call establishment delay is the summation of time taken by network to locate MT when an incoming call arrives for it and the time taken to assign the resources to handle that call. Generally it is assumed that the time taken to assign resources is constant therefore the focus should be on minimizing the time taken by the network to locate the target MT. Hence, the above three parameters are related to each other as follow:

$$Cost_{(Update)} \; \alpha \; [1/Cost_{(Paging)}]$$

$$Cost_{(Paging)} \; \alpha \; Paging \; area$$

$$Call \; establishment \; delay \; \alpha \; number \; of \; paging\text{-}miss$$

D. *Involvement of Mobile Terminal:* The MT has limited battery power; greater involvement of MT in a LM scheme means higher battery consumption on MT side.

E. *Computational overhead:* A LM scheme is not preferred if it requires high computational capabilities. This is the reason for the wide acceptance of simple static LM schemes over high performance dynamic LM schemes.

$$Computational \; complexity \; \alpha \; [1/acceptance \; of \; LM \; Scheme]$$

Vijendra Singh Bhadauria et al. / International Journal on Computer Science and Engineering (IJCSE)

The biggest challenge in location management is to provide good trade-off between location update and paging operations. It must be assured that if one of them goes down then the other should not rise significantly. In future, LPCIC rule (L-LU cost, P-Paging cost, C-Call establishment delay, I-Involvement of MT, C-Computational overhead) can be used for rating the LM schemes according to the number of parameters they minimize.

## IX. LOCATION MANAGEMENT ARCHITECTURE IN 3-G NETWORK

In 3G, Gateway location register (GLR) is the additional database as compared to 2G network, as shown in Fig.3. The subscriber profile information is copied from the HLR to GLR so that the GLR can handle the LU messages from VLRs as if it is the HLR of the subscribers. It handles the location update procedure locally for the movement within the visited network, thus reducing the costly inter-visited network signaling. For VLR at the visited network, GLR is treated as the roaming user's HLR located at the visited network whereas for HLR at the home network, GLR is treated as the VLR.

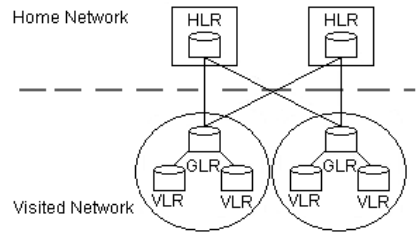

Figure 3. Location management architecture in 3-G network

## X. CONCLUSION

The purpose of this article is to present a comprehensive classification and comparative study of major LM schemes in cellular mobile networks. It focuses on major drawbacks of existing LM schemes and parameters influencing the performance of a LM scheme. The main contribution of this paper is the guideline for the development of a location management scheme that would probably be helpful to the community related to this field.

## AUTHORS PROFILE

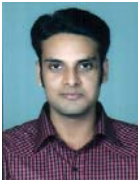

**Mr. Vijendra Singh Bhadauria** is M.Tech. in Computer Science and engineering. Presently he is pursuing Ph.D. in same theme from RGTU, (Bhopal) India. He is IBM certified Database Associate and his fields of interest include Mobile computing, Data Structures and Algorithms, Programming Logics and Network Security. Currently he is working as Assistant Professor in Computer science and engineering department of SIRTE, Bhopal (India).

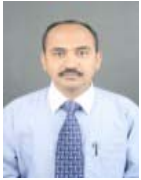

**Dr. Sanjeev Sharma** has graduated in Electrical & Electronics in 1991 from Samrat Ashok Technical Institute, India and post graduated in Microwave and Millimeter from Maulana Azad College of Technology. He completed his Doctorate in Information Technology from Rajiv Gandhi Proudyogiki Vishwavidyalaya. Currently he is working as head in School of IT, RGTU (Bhopal), India. He possesses teaching and research experience of more than 15 years. His areas of interest are Mobile Computing, Data Mining and Information Security. He has edited proceedings of several national and international conferences and published more than 25 research papers in reputed journals.

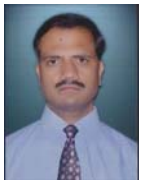

**Dr. Ravindra Patel**, Reader and Head, Department of Computer Applications at Rajiv Gandhi Technological University, (Bhopal) India. He is Ph.D. in Computer Science. He possesses more than 10 years of teaching experience in post-graduation. He has published more than 15 research papers in international and national journals, and conference proceedings. He is member of ISTE and International Association of Computer Science and Information Technology (IACSIT).